\documentclass[conference]{IEEEtran}
\IEEEoverridecommandlockouts
\usepackage{cite}
\usepackage{amsmath,amssymb,amsfonts}
\usepackage{algorithmic}
\usepackage{graphicx}
\usepackage{textcomp}
\usepackage{xcolor}
\usepackage[nolist]{acronym}
\usepackage{paralist}

\begin{document}

\makeatletter
\def\ps@IEEEtitlepagestyle{%
\def\@oddfoot{\parbox{\textwidth}{\footnotesize
Author's version of a paper accepted for publication in Proceedings of the 2021 International Conference on Smart Energy Systems and Technologies (SEST). 
\\
\textcopyright{} 2021 IEEE. 
Personal use of this material is permitted.  
Permission from IEEE must be obtained for all other uses, in any current or future media, including reprinting/republishing this material for advertising or promotional purposes, creating new collective works, for resale or redistribution to servers or lists, or reuse of any copyrighted component of this work in other works.\vspace{1.2em}}
}%
}
\makeatother

\begin{acronym}
\acro{sg}[SG]{smart grid}
\acroplural{sg}[SGs]{smart grids}
\acro{der}[DER]{distributed energy resource}
\acroplural{der}[DERs]{distributed energy resources}
\acro{ict}[ICT]{information and communication technology}
\acro{fdi}[FDI]{false data injection}
\acro{scada}[SCADA]{Supervisory Control and Data Acquisition}
\acro{mtu}[MTU]{Master Terminal Unit}
\acroplural{mtu}[MTUs]{Master Terminal Units}
\acro{hmi}[HMI]{Human Machine Interface}
\acro{plc}[PLC]{Programmable Logic Controller}
\acroplural{plc}[PLCs]{Programmable Logic Controllers}
\acro{ied}[IED]{Intelligent Electronic Device}
\acroplural{ied}[IEDs]{Intelligent Electronic Devices}
\acro{rtu}[RTU]{Remote Terminal Unit}
\acroplural{rtu}[RTUs]{Remote Terminal Units}
\acro{iec104}[IEC-104]{IEC 60870-5-104}
\acro{apdu}[APDU]{Application Protocol Data Unit}
\acro{apci}[APCI]{Application Protocol Control Information}
\acro{asdu}[ASDU]{Application Service Data Unit}
\acro{io}[IO]{information object}
\acroplural{io}[IOs]{information objects}
\acro{cot}[COT]{cause of transmission}
\acro{mitm}[MITM]{Man-in-the-Middle}
\acro{fdi}[FDI]{False Data Injection}
\acro{ids}[IDS]{intrusion detection system}
\acroplural{ids}[IDSs]{intrusion detection systems}
\acro{siem}[SIEM]{Security Information and Event Management}
\acro{mv}[MV]{medium voltage}
\acro{lv}[LV]{low voltage}
\acro{cdss}[CDSS]{controllable distribution secondary substation}
\acro{bss}[BSS]{battery storage system}
\acroplural{bss}[BSSs]{battery storage systems}
\acro{pv}[PV]{photovoltaic inverter}
\acro{mp}[MP]{measuring point}
\acroplural{mp}[MPs]{measuring points}
\acro{dsc}[DSC]{Dummy SCADA Client}
\acro{fcli}[FCLI]{Fronius CL inverter}
\acro{fipi}[FIPI]{Fronius IG+ inverter}
\acro{sii}[SII]{Sunny Island inverter}
\acro{tls}[TLS]{Transport Layer Security}
\acro{actcon}[ActCon]{Activation Confirmation}
\acro{actterm}[ActTerm]{Activation Termination}
\acro{rtt}[RTT]{Round Trip Time}
\acro{c2}[C2]{Command and Control}
\acro{dst}[DST]{Dempster Shafer Theory}
\acro{ec}[EC]{Event Correlator}
\acro{sc}[SC]{Strategy Correlator}
\acro{ioc}[IoC]{Indicator of Compromise}
\acroplural{ioc}[IoCs]{Indicators of Compromise}
\acro{ot}[OT]{Operational Technology}
\end{acronym}

\title{Towards an Approach to Contextual Detection of Multi-Stage Cyber Attacks in Smart Grids}

\author{
\IEEEauthorblockN{%
Ömer Sen\IEEEauthorrefmark{1},
Dennis van der Velde\IEEEauthorrefmark{1},
Katharina A. Wehrmeister\IEEEauthorrefmark{1},
Immanuel Hacker\IEEEauthorrefmark{1},
Martin Henze\IEEEauthorrefmark{3},
Michael Andres\IEEEauthorrefmark{1}}

\IEEEauthorblockA{%
\IEEEauthorrefmark{1}\textit{Digital Energy, Fraunhofer FIT,} Aachen, Germany\\
Email: \{oemer.sen, dennis.van.der.velde, immanuel.hacker, katharina.wehrmeister, michael.andres\}@fit.fraunhofer.de}

\IEEEauthorblockA{%
\IEEEauthorrefmark{3}\textit{Cyber Analysis \& Defense, Fraunhofer FKIE,} Wachtberg, Germany\\
Email: martin.henze@fkie.fraunhofer.de}
}

\maketitle

\begin{abstract}
Electric power grids are at risk of being compromised by high-impact cyber-security threats such as coordinated, timed attacks.
Navigating this new threat landscape requires a deep understanding of the potential risks and complex attack processes in energy information systems, which in turn demands an unmanageable manual effort to timely process a large amount of cross-domain information. 
To provide an adequate basis to contextually assess and understand the situation of smart grids in case of coordinated cyber-attacks, we need a systematic and coherent approach to identify cyber incidents. %
In this paper, we present an approach that collects and correlates cross-domain cyber threat information to detect multi-stage cyber-attacks in energy information systems.
We investigate the applicability and performance of the presented correlation approach and discuss the results to highlight challenges in domain-specific detection mechanisms.
\end{abstract}


\section{Introduction} \label{sec:introduction}
Power grids are currently undergoing far-reaching changes and evolving into \acp{sg} to accommodate the increasing penetration by \acp{der}~\cite{1_tuballa2016review}.
Intelligently integrating the actions of all connected stakeholders using \ac{ict} requires the secure and controllable integration of volatile \acp{der} as well as novel grid components such as heat pumps and electric vehicles, which \acp{sg} can provide a foundation for~\cite{39_smartrgird2020book}.
However, the increasing prevalence of \ac{ict} and the convergence between information and operational technology in the energy sector means that more access points to control systems are emerging and, consequently, new cybersecurity challenges are arising~\cite{3_van2020methods,krause2021survey,serror_iiot_2021}.
This new threat landscape poses risks of incidents with critical disruptive consequences to grid operation~\cite{56_kimani2019cyber}.
Here, cyber countermeasures such as \acp{ids} can help identify early indicators of an attack and provide an information base for deriving appropriate response and mitigation measures~\cite{9_mendel2017smart}.
Detecting intrusions by unauthorized persons into the central monitoring and control system of network operators, especially attacks within the network perimeter, is fraught with challenges.
For example, commands with potentially negative effects on the grid may originate from legitimate but compromised hosts.
Consequently, it is not sufficient to secure and monitor communication paths.
This requires the contextual correlation of indicators of an attack from different components and temporal developments that unfold over time~\cite{40_chromik2019process, 41_liu2016situational,sen2021replicating}.
Common approaches to address this challenge are based on \ac{siem} systems that aggregate and correlate from different \acp{ids} to provide real-time traffic analysis, early detection of attack-related events, and event correlation.
However, process networks in power grids offer special advantages in detecting implausible events and anomalies in the process environment due to the static network structure, deterministic data traffic, and physically constrained process information~\cite{klaer_sgam_2020}. %

To remedy security issues in \acp{sg}, different streams of research address the challenges of automatically analyzing large amounts of cyber threat data.
In particular, an approach of physical consistency checking at the substation level has been proposed to validate process data according to a set of constraints, thus noticing when an individual substation enters a ``bad state'' that represents, e.g., a physical instability~\cite{40_chromik2019process}.
Further research aimed to reduce the false positive rate of rule-based \ac{ids} solutions by correlating different \ac{ids} events to create attack scenarios and using machine learning to teach a system which attacks reported by an \ac{ids} is likely to be genuine~\cite{44_zomlot2014handling}.
Using a situational awareness approach where sensors distributed in the \ac{sg} relay relevant information to a command center (similar to a \ac{siem} system), event correlation and integrity checking can be used to detect complex attacks~\cite{53_mavridou2011situational}.
Many of the related works present approaches for contextual assessment and reconstruction of security incidents in \acp{sg}.
However, limiting knowledge acquisition to events from the same source and not considering alarms from other security systems or logs from other \ac{ict} network components excludes additional information from different perspectives.
This limits the extent to which a potential incident may be understood and assessed.
Thus, when detecting complex attacks with data from multiple sources, additional information such as domain-specific knowledge from power grids enriches the detection.
For example, process data in the form of data points, the flow of data in the \ac{ot} environment, the \ac{ict} network topology, and the interaction between assets provide an additional perspective for a holistic and global view of the cyber-physical situation.

To provide a foundation for detecting and preventing such attacks, this paper addresses the detection of multi-stage cyber-attacks by leveraging domain-specific attributes of attack indicators within a context-based, cross-domain correlation approach of \ac{ict} security incident indicators.
To this end, we propose a \ac{siem}-based \textbf{d}etection system \textbf{o}f \textbf{m}ulti-stage \textbf{c}oordinated \textbf{a}ttacks (DOMCA) to identify the appropriate attack evolution and strategy.
Our contributions are:
\begin{enumerate}
    \item We propose an event correlation mechanism to identify complex attack actions based on cyber threat observations (Section~\ref{subsec:framework_eventcorr}).
    \item We present and describe a structured approach to detect strategies of multi-stage attacks in energy information systems (Section~\ref{subsec:framework_strategycorr}).
    \item We demonstrate and discuss the performance of our proposed framework against different attack scenarios in a simulation environment (Section~\ref{sec:result}).
\end{enumerate}

\section{Cyber Security in Smart Grids} \label{sec:background}
In this section, we set the foundation for our work by providing a brief overview of cyber-security issues in process networks, as well as detection and correlation mechanisms for identifying security incidents.

\subsection{Cyber Security \& Power Grids} \label{subsec:background_cybersec}
The integration of \ac{ict} into power grids enables the exchange of process data, e.g., measurement values from sensors, via \acp{rtu} to \ac{mtu} within \ac{scada} systems~\cite{18_radoglou2019attacking}. %
The \ac{scada} system is responsible for monitoring received data and issuing alarms in case of disturbances to the grid (e.g., critical load, voltage threshold violations, power quality disturbances)~\cite{39_smartrgird2020book}.
Based on the higher-level decision and optimization functions, such as optimal power flow calculations, suitable commands that control the actuators via the field devices in the process network are determined~\cite{40_chromik2019process}. %
This process is performed to optimize the grid state considering stability, resource utilization, and flexibility constraints.
Traditional process networks were characterized by isolated, proprietary, legacy components that created a barrier to unauthorized third parties.
This has been dismantled by the increasing integration of \ac{ict} and the interconnection of various grid assets and actors, leading to the emergence of a new cyber threat landscape~\cite{42a_nazir2017assessing}.
The new access points, traversable communication paths, and vulnerabilities can be leveraged in coordinated cyber-attacks that aim to disrupt or damage the power grid by intercepting, manipulating, and spoofing communications between its \ac{scada} components on a large scale~\cite{43_eder2017cyber}.
For example, the Stuxnet attack in 2010 reportedly severely disrupted Iran's nuclear program~\cite{55_kshetri2017hacking}.
Further, between 2013 and 2014, a Stuxnet-like Trojan called Havex compromised the control systems of more than 1,000 energy companies in 84 countries~\cite{55_kshetri2017hacking}.
Moreover, coordinated attacks in 2015 and 2016 in Ukraine led to a temporary power outage affecting more than 200,000 customers~\cite{55_kshetri2017hacking}.

\subsection{Contextual Detection of Cyber Incidents} \label{subsec:background_siem}

To timely detect coordinated cyber-attacks, \ac{ids} solutions automate the process of intrusion detection by recognizing either attack indicators based on the normal operation (anomaly-based) or attack signatures (misuse-based) or their combined knowledge~\cite{44_zomlot2014handling}.
Traditional \acp{ids} monitor only the \ac{ict} network and/or its host components (e.g., login attempts, network scans, suspicious log traffic, or syslog) without involving the process semantics of the power grid~\cite{45a_chromik2017context}.
Contextual detection can be achieved based on a \ac{siem} system that combines functionalities such as security data collection and consolidation, long-term data storage, automation of analysis and reporting, and real-time monitoring and correlation of events from various data sources~\cite{46a_vielberth2021security}.
Data aggregation involves collecting log and event data from different types of sources (e.g., \ac{ids} or firewalls)~\cite{48a_bryant2017novel}.
It also involves normalizing data to a common format as well as synchronizing associated event fields such as timestamps, providing comparable and accessible characteristics of the data for processing and correlation~\cite{47a_radoglou2021spear}.
In particular, correlation and reasoning approaches that aim to classify and infer characteristics and relationships between entities involved in multi-stage attacks use contextual information in the graph-based representation of such attacks.
To this end, attack graphs have proven beneficial to model the hierarchical unidirectional dependency between the steps within a multi-stage attack and their transitions~\cite{50a_angelini2018attack}.
By having nodes with multiple successors or predecessors, attack graphs can represent strategies that involve multiple possible paths to an attacker's target~\cite{50a_angelini2018attack}. %
Alternatively, the Kill-Chain modeling concept provides an approach for structuring multi-stage attacks aimed at disrupting or destroying vital processes or devices.
Steps within the structure include gaining access to and information about the target system, developing and testing new capabilities on the compromised targets, exploiting vulnerabilities and moving laterally in the network, building \ac{c2} infrastructure, and acting on the objection (e.g., disrupting grid operations)~\cite{52a_kour2020railway}.

\section{Multi-Staged Attack Detection System} \label{sec:framework}
The correlation of cyber threat information and process data faces challenges, such as accounting for false positives that occur in traditional probabilistic-based correlation approaches~\cite{44_zomlot2014handling}.
The approaches assign simple probability values to statements about an attack but do not provide a representation for the certainty of those assignments~\cite{44_zomlot2014handling}.
Thus, an appropriate quantification method is needed to model the level of confidence of detected attack indicators.
This challenge is exacerbated by the lack of data to quantify the likelihood of an attack, particularly attack data from critical infrastructure. %
Subsequently, assigning probabilities to indicators of an attack a priori becomes infeasible~\cite{44_zomlot2014handling}.
To address this issue, theories that deal with epistemic uncertainty can be used, such as \ac{dst}~\cite{51_sentz2002combination}.
\ac{dst} is seen as a generalization of traditional Bayesian probability theory, making it possible to assign a probability to sets of statements rather than individuals.
This allows the combination of evidence from multiple sources without a priori knowledge, i.e., a priori probability distributions, about system states~\cite{51_sentz2002combination}.
In the following, we present the architecture of DOMCA to detect the corresponding attack evolution and strategy based on domain-specific attribution and contextual correlation of cyber incident indicators using \ac{dst}.

\subsection{Framework Overview} \label{subsec:framework_overview}
\begin{figure}
    \centerline{\includegraphics[width=\columnwidth]{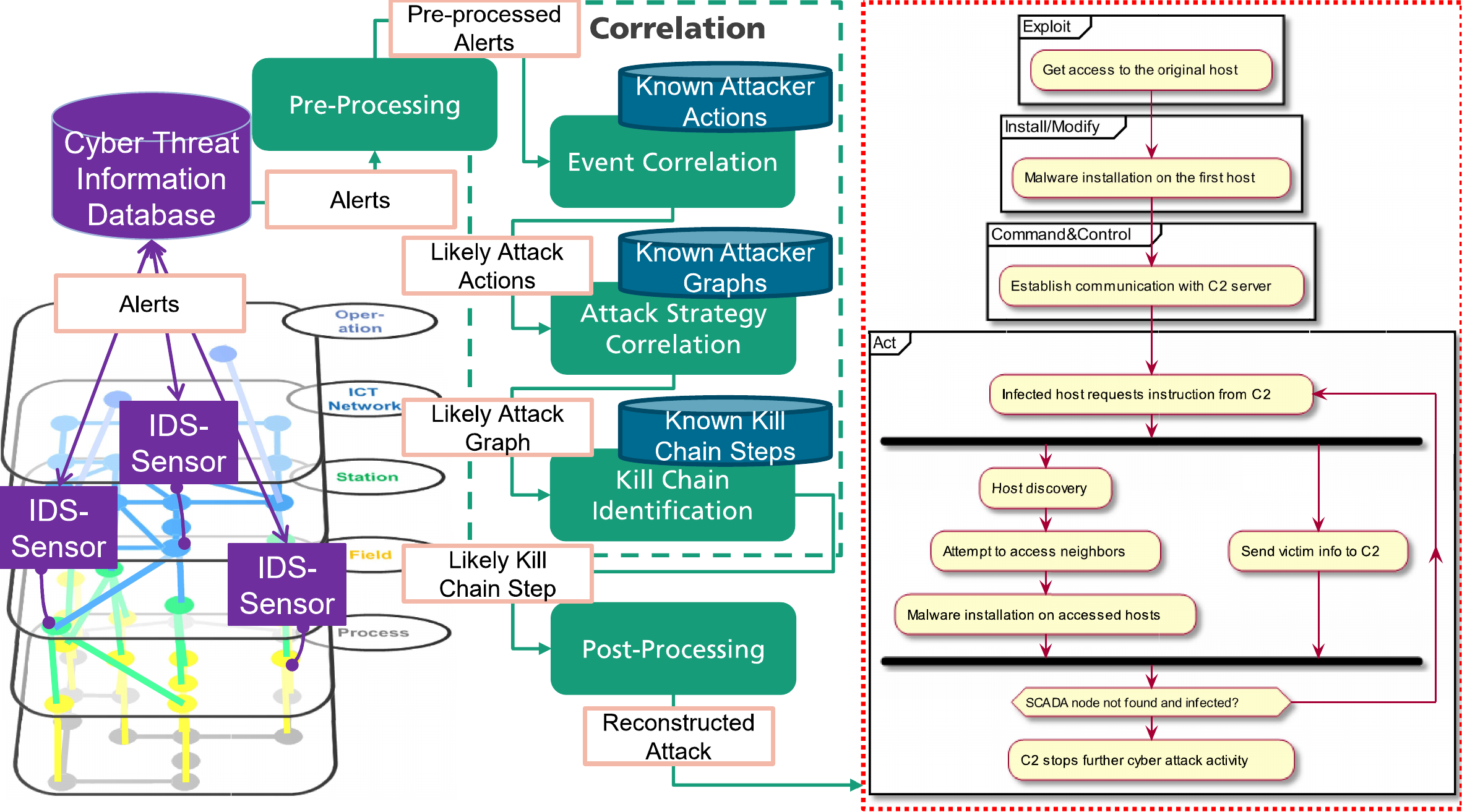}}
    \caption{Structural overview of the presented kill-chain-based correlation and detection system for contextual detection of multistage cyber incidents.}
    \label{fig:framework_overview}
    \vspace{-1em}
\end{figure}
Our core idea is to reconstruct the propagation of a cyber-attack in several stages and identify corresponding appropriate strategies, as shown in Figure~\ref{fig:framework_overview}.
To reconstruct cyber incidents based on attack indicators, DOMCA's \textit{pre-processing} component takes domain-specific process, communication, and semantic indicators captured by distributed sensors.
Further, DOMCA pre-processes data into normalized alarms for effective analysis (cf. Section~\ref{subsec:framework_preprocessing}).
For simplicity, we represent various monitoring and attack indicator generation from multiple sources as distributed \ac{ids} sensors.
In addition, we envision a central architectural framework (e.g., at the operations center level) to increase situational awareness at a more global level.
Although outside the scope of this paper, architectural security can be enhanced by a distributed ledger communication layer connecting the sensors and the central correlation framework.
Given a set of known possible actions that an attacker could perform, the \textit{\ac{ec}} uses the pre-processed attack indicators to determine their possible occurrences, assigning each a confidence level via \ac{dst} (cf. Section~\ref{subsec:framework_eventcorr}).
Once the event correlation process has identified all potentially performed attack actions, the \textit{\ac{sc}} uses \ac{dst} and custom combination rules to determine possible paths through known attack graphs.
By considering both the mass of each detected action and masses assigned to attack graph edges, the \ac{sc} therefore identifies feasible attack strategies based on current observations (cf. Section~\ref{subsec:framework_strategycorr}).
The final step of the correlation process is \textit{kill-chain identification}, which is responsible for determining the most likely attack path and corresponding graph based on \ac{sc} results.
Subsequently, this determines the kill-chain step corresponding to the last step of the chosen path (cf. Section~\ref{subsec:framework_killchain}). %
Finally, the \textit{post-processing} component is responsible for the comprehensive visualization and higher-level processing of the result (cf. Section~\ref{subsec:framework_preprocessing}).

\subsection{Pre-Processing} \label{subsec:framework_preprocessing}
The preprocessing component of our framework brings diverse input information from different data sources into a comparable and processable output format via normalization processes.
\begin{table}[h!]
\centering\small
\caption{Domain-specific attribution of alarms within events.}
\begin{tabular}{||p{2.3cm} p{5.5cm}||}
 \hline
 Fields & Description \\
 \hline\hline
 \verb|IoC| & Participation in an attempt to access a host.\\
 \verb|ADR_FROM_CHECK| & Suspicious source of the message.\\
 \verb|ADR_TO_CHECK| & Suspicious destination of the message.\\
 \verb|CON_CHECK| & The connection over which the packet was sent is not allowed.\\
 \verb|DP_FROM_CHECK| & The packet contains data points that are unexpected for the source host.\\
 \verb|DP_TO_CHECK| & The packet contains data points that are unexpected for the receiver host.\\
 \verb|CYCLE_CHECK| & A message that normally arrives cyclically deviates from its schedule.\\
 \hline
\end{tabular}
\label{tab:framework_preprocessing}
\end{table}
A predominant source of input is made up of alerts from distributed \ac{ids} sensors in the process network, which in our work are conceptualized as specification-based \ac{ids}.
The sensors occupy selected \ac{ict} network edges and perform domain-specific attribution of the captured packets (cf. Table~\ref{tab:framework_preprocessing}). %
In case of a failed check, a packet event is generated containing the name of the failed criterion, along with information about the packet.
This includes its payload, payload type (e.g., a command, measurement, monitoring, IT payload), a timestamp of the alarm's occurrence, the source and destination addresses of the packet, the endpoints of the monitored \ac{ict} edge, and the ID of the sensor reporting the packet.
Furthermore, for identifying the consistent path of a message through multiple hosts (IT and OT components), the chronological ordering of packet events is performed based on timestamps and the topological relationship of the monitored \ac{ict} edges.
Sensor placement is an important factor in this context, however, some missing sensors can be accounted for by identifying pairs of paths characterized by consistent connectivity in the network, sensor coverage, and marginal timing differences of events between pairs.
Each pair found in this way is consolidated into a new path that combines the information of the stored pairs, including the concatenation of their host paths and the consolidation of possible failed security checks.
The final format of the normalized events is clustered by the constructed paths in (cf. Table~\ref{tab:framework_normalized}).
\begin{table}[h!]
\centering
\caption{Normalized output format of the pre-processing component.}
\begin{tabular}{||p{1.7cm} p{6cm}||}
 \hline
 Fields & Description \\
 \hline\hline
 \verb|EVENT_ID| & List of event IDs assigned to this message path.\\
 \verb|SEND_TIME| & Sending time of the original responsible host.\\
 \verb|RECEIVE_TIME| & Receiving time of the last destination host.\\
 \verb|FROM_HOST| & Host responsible for the message.\\
 \verb|PASSED_HOSTS| & Hosts that the message passed through on its way.\\
 \verb|TO_HOST| & Final destination of the message.\\
 \hline
\end{tabular}
\label{tab:framework_normalized}
\end{table}

\subsection{Event Correlator} \label{subsec:framework_eventcorr}
The objective of the \ac{ec} is to use normalized alarms to draw conclusions about an attacker's actions and sort the identified actions in chronological order.
Subsequently, the \ac{sc} can use this information to identify attack strategies that use a subset of the known action set.
First, the \ac{ec} performs an analysis of suspicious communications to identify infected hosts based on frequently sent suspicious packets. %
Additionally, the \ac{ec} attempts to identify the position of the \ac{c2} coordinator. %
This is done based on the structure of the infected hosts' communications, assuming that the node with the most outgoing suspicious messages is the \ac{c2} host.
After this analysis, the \ac{ec} detects individual actions based on normalized alerts, such as an attacker's attempt to access a host (e.g., \ac{rtu}). %
To identify such an access attempt, it first generates a list of all occurrences of access attempt indicators (e.g., network scan, attempted or suspicious login, privilege escalation).
Then, it sorts them in chronological order, grouping them by the \ac{ict} network host they targeted, and consolidating them across all detected \acp{ioc} into pairs of source and destination hosts.
Afterward, the \ac{ec} assigns mass functions to each possible access attempt.
These weigh the relevance of the observed \acp{ioc} in the context of access attempts, taking into consideration their types, chronological order, and possible associations with other related access attempts \acp{ioc}.
Based on the access attempt detection action, network access attempt detection is performed by iterating the list of source-destination pairs to determine when each infected host first attempted to infiltrate another.
Also, malware installation detection on likely infected \ac{ict} hosts is performed (e.g., compromised \acp{rtu}), using the timestamp of the last detected access attempt on such a host along with the first detected suspicious message sent from it. %
After malware installation, an infected node may not immediately have a preconfigured initial task and instead send a request to the \ac{c2} host asking for instructions (e.g., \acp{rtu} waiting for control action on data manipulation).
The \ac{ec} can detect this by recognizing that such a message was sent outside of specified allowed connections (e.g., server outside the process network perimeter), and by verifying that the \ac{c2} host has sent a message back to the requestor within a certain time.
Detection of communication-dependent attacker actions using compromised hosts (e.g. to collect information on the infiltrated \ac{ict} network) considers any normalized alert reporting suspicious communication.
Particularly, alert that is not an access attempt or command request, and checks for \ac{c2} network exclusion, an indication of bilateral communication (e.g., horizontal communication between \acp{rtu}). %
The \ac{ec} uses \ac{dst} to store the confidence in detected attack actions.
Specifically, Zhang's combination rule~\cite{51_sentz2002combination} is used to combine mass distributions of different statements as additive evidence.
Additionally, if multiple actions are reliant on the same alert, their current mass will each be combined with an ``impact mass'' that represents the negative impact that this has on their legitimacy (e.g., indication of legitimate but compromised \acp{rtu}).
The \ac{ec} finally outputs a list of attacker actions with associated confidence values, timestamps, and affected hosts for a preconfigured time horizon.

\subsection{Strategy Correlator} \label{subsec:framework_strategycorr}
\begin{figure}
    \centerline{\includegraphics[width=\columnwidth]{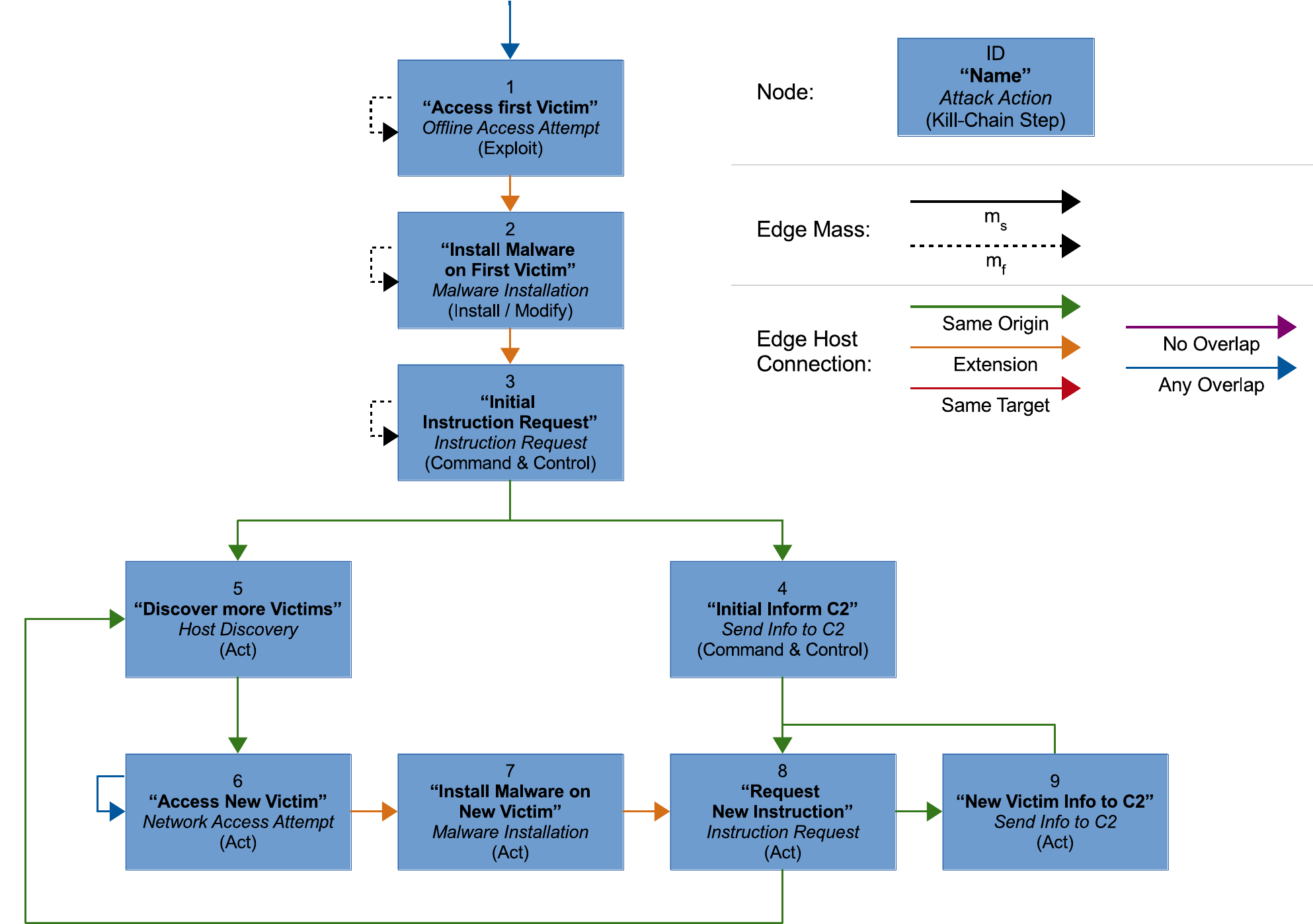}}
    \caption{Exemplary attack graph based on attacker actions for the Havex attack. Edges represent mass assignments and inference links between actions.} %
    \label{fig:sc_havexag}
    \vspace{-1em}
\end{figure}
Based on the set of attacker actions determined by the \ac{ec}, the \ac{sc} begins the task of identifying which known attack strategies fit the observed behavior.
Using a set of predefined attack graphs, the likely attack incident and possible attack evolution are determined (cf. Figure~\ref{fig:sc_havexag}).
Our concept of an attack graph has structural similarities to exploit dependency graphs~\cite{50a_angelini2018attack}.
However, we focus on general attack actions rather than the exploitation of specific vulnerabilities.
Therefore, attack actions can represent different types of steps that an attacker can take in different domains and situations.%
Thus, a node in an attack graph contains a unique identifier within the attack graph, a description that links the action to the overall strategy, the attack action represented by that node, and a phase within the kill chain to which that step of the attack strategy belongs.
Nodes can have one or more predecessors and successors representing decisions an attacker can make within the strategy, as well as consequences caused by the state of the \ac{ict} network or the attacker.
The edges in the attack graph represent the transition between actions, including possible connections between the hosts involved.
Also, each edge contains a mass distribution, which depends on the probability of the connected actions succeeding each other within the represented attack strategy.
Thus, an attack graph consists of multiple nodes and edges that form paths representing a sequential attack process.
The attack graph in its essence represents the attack strategy defined with the focus on \acp{sg}.
Based on a predefined set of such attack graphs, and after receiving the attacker actions from \ac{ec}, \ac{sc} starts its analysis.
The initial process involves adding new edges to each attack graph, taking into account possible undetected actions or irregularities in the attacker's behavior.
After the graphs are prepared, the attacker's possible paths through each known attack graph are reconstructed, each resulting in a chronological list of traversed nodes and the hosts involved.
Furthermore, overall mass distribution is assigned to each path, taking into account both the masses of the traversed edges and considered actions.
Beforehand, action masses have been adjusted to contain a relatively high uncertainty.
This is necessary because of the additive nature of Zhang's combination rule, which would result in very high certainty with few considered actions otherwise. %
The attack graphs themselves are also assigned mass distributions.
These depend on the overlap between detected actions and those contained in each attack graph, as well as the mass of the path with the highest belief value running through the graph.
It is then checked whether the attack path with the highest belief value is contained in the attack graph with the highest belief value and whether they both exceed their respective confidence thresholds. %
At this stage, the \ac{sc} finishes its analysis and outputs a collection of pairs containing reconstructed attack paths and associated graphs with their corresponding belief values for the next component to consider.

\subsection{Kill-Chain Identification} \label{subsec:framework_killchain}
After the correlation performed by \ac{sc}, the Kill-Chain identification component must decide whether an attacker is present in the environment and if so, which known graph most closely represents the attacker's behavior. %
To determine the most credible and plausible attack graph and path pair within the provided set, a successive comparison of the mass distribution of each pair is performed against a predefined threshold and cutoff values.
The corresponding plausibility and belief values of the attack graph and path are checked to see if they exceed the lower predefined threshold values representing the cutoff process.
After determining the most credible attack path and graph, the system checks if the path passes through the graph and outputs it as the optimal solution.
The name of the attack strategy corresponding to the attack graph identified by this process is output as the detected attack strategy.
Furthermore, the last kill-chain phase of the attack is the kill-chain phase stored in the last node reached by the path.
This identifies the attacker's current phase within a kill-chain-based process, indicating what state the attacker is currently in.
If no matching pair of path and graph could be determined, either no attack occurred or an attack that did not match any of the known strategies occurred.
Based on the correlation results, the system can identify whether an attack occurred within a certain time horizon (meaningful output available), how the attack evolved (detected attack path), and what strategy the attacker followed (detected attack graph).
Moreover, it identifies which attack phase was last observed (detected kill-chain phases), and which host was involved in the attack process (list of infected hosts).

\subsection{Post-Processing} \label{subsec:framework_postprocessing}
Upon a successful correlation process, a post-processing component can be used for further higher-level processing and visualization of identified attack graph, path, and actions.
Streamlined and visualized correlation results could make cyber incidents understandable to a user and be presented with the appropriate confidence level, for example in security operations centers, for potential incident response.
Additionally, post-processing can also be part of a decision support system in the incident response task area to automate and support containment and mitigation strategies by also predicting the next step of attacker actions based on correlation results.

\section{Evaluation \& Discussion} \label{sec:result}
In the following, we evaluate and discuss the performance of DOMCA concerning the reconstruction of multi-staged attacks within a simulation environment according to~\cite{klaer_sgam_2020}.

\subsection{Procedure for the Investigation} \label{subsec:result_proecdure}

For the investigation, attack scenarios are simulated in an \ac{sg} simulation environment according to~\cite{klaer_sgam_2020} with different strategies, each representing a Havex, Stuxnet, randomized (performing random attack actions), or no attack incident.
The network parameters that can be modified to include the vulnerability of individual hosts, the configuration of hosts with no vulnerability (no successful access attempts), and hosts that are explicitly vulnerable to remote access attempts.
Besides, sensor placement is an important aspect of parametrization.
It affects the functionality of DOMCA by influencing the observation provided, i.e., directly affecting situational awareness. This especially applies to sensors near the \ac{c2} host.
As part of the investigation, we performed a total of 207 simulation runs with approximately evenly distributed runs of the attack scenarios presented in the environment.

\subsection{Classification Accuracy Evaluation} \label{subsec:result_res}
\begin{figure}
    \centerline{\includegraphics[width=\columnwidth]{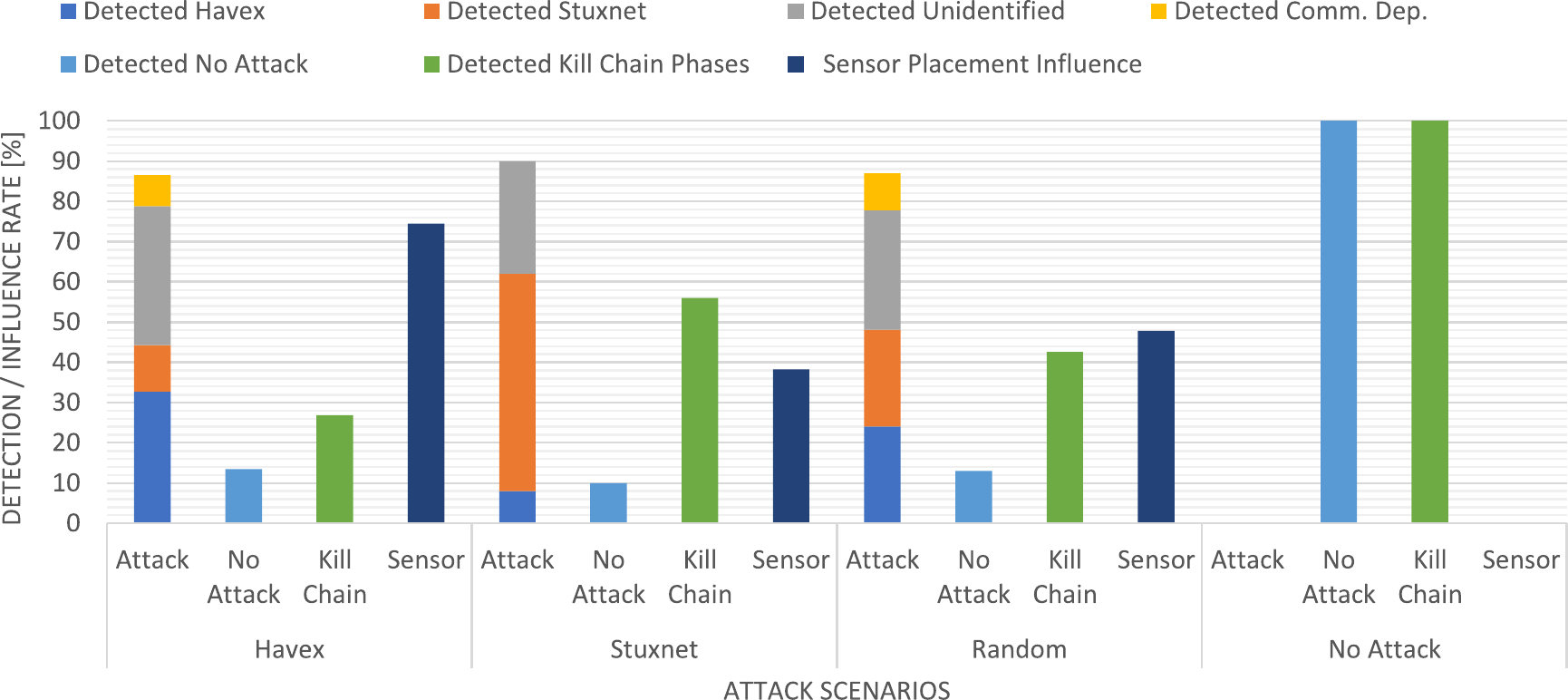}}
    \caption{The classification accuracy assessment chart shows on the x-axis the attack scenarios performed, including the ``no attack'' event, and on the y-axis the distribution of the detection rate of attack strategies, kill-chain phases, and the influence of sensor placement on detection quality.}
    \label{fig:result_acc}
    \vspace{-1em}
\end{figure}
Figure~\ref{fig:result_acc} illustrates for each real attack scenario the detection rate of attacker presence, Kill-Chain step, as well as the distribution of detected strategies among simulation runs.
The plot also illustrates the average impact of sensor placement on classification accuracy.
In general, no false positives were observed in the experiments, which means that the system never detected an attacker when none was present.
The average detection rate concerning the presence of any attacker was 87.86\%. %
In a ``randomized'' attack, where random attack actions are performed, the attack is still detected in the vast majority of cases.
However, it is often incorrectly assigned to a known attack graph. %
When examining the detection rate of the system in identifying infected hosts and individual attack actions, we found that its accuracy is affected by the placement of a sensor in the \ac{c2} node's communications.
Finally, our results regarding the correct determination of the last kill chain phase of an attacker shows a detection accuracy of 56.38\% for different kill chain phases, regardless of the chosen strategy.

\subsection{Discussion} \label{subsec:result_dis}
As our evaluation shows, our proposed approach DOMCA reliably detects the presence of an attacker in an \ac{ict} network, with no false positives observed in our experiments. %
We also found that the detection accuracy of individual strategies depends on several factors, e.g., placement of the sensors close to the \ac{c2} nodes. %
In particular, the presence of a sensor monitoring communication with a \ac{c2} host contributes significantly to the accuracy of detecting the strategy used by an attacker.
Additionally, the duration of a simulation and corresponding attack has a large impact on the effectiveness of the system by affecting the false-negative rate that occurs for a given duration.
This can result in an attack being detected in its initial stages but not correctly mapped to a known attack graph.
The system might have benefited from additional functionality to detect these early, more universal kill chain phases independent of a particular strategy.
However, when the correct strategy was first identified, the accuracy of detected kill chain phases was consistently above 97\%, as well as when kill chain phases were correctly not detected, such as in the ``no attack" scenario. %
Overall, depending on the observable network area and the previous knowledge of attack actions and strategies, DOMCA can reliably reconstruct the attack evolution process and provides an advanced basis for attack prediction and mitigation.

\section{Conclusion} \label{sec:conclusion}
Detecting and defending against increasingly complex cyber-attacks requires an approach that enables an understanding of the current cyber-physical situation, especially in the context of communication-dependent processes.
To this end, in this paper, we present a kill-chain-based correlation approach - DOMCA - to contextually identify multi-stage cyber-attacks with severe consequences for reliable power supply in \acp{sg}. %
We discuss the design and subsequent implementation of DOMCA, which consists of a data formatting normalizer, an attack action, and a strategy correlator respectively, as well as a Kill-Chain identifier responsible for identifying the attacker's most likely strategy and the current kill-chain stage. %
Furthermore, we evaluate DOMCA's detection rate against different attack scenarios and parametrized network settings. %
Our key findings are that DOMCA can reliably detect an attacker in the simulated energy \ac{ict} environment for our conducted attack scenario experiments.
Notably, the accuracy of the kill-chain phase and attack strategy identification is highly dependent on the placement of sensors, the extent of observation, and the degree of attack development.
Future work includes further investigation of the applicability of DOMCA in a realistic \ac{sg} environment and other use cases (e.g., local energy communities, microgrids) to draw reliable conclusions about the effectiveness of the proposed approach.
In addition, secure-by-design principles will be explored with respect to the architecturally central framework using communication layer security technologies such as distributed ledgers.
Nevertheless, even in its current form, DOMCA can reliably reconstruct the development process and strategy of known attacks and provide an advanced basis for future research in decision support systems for actions to mitigate such attacks.
In addition to its applicability in \acp{sg}, DOMCA can be extended to other critical infrastructures if attack graphs and actions as well as domain-specific attribution are adapted.

\noindent\textsc{Acknowledgments}\hspace{1em}
This work has partly been funded by the German Federal Ministry for Economic Affairs and Energy (BMWi) under project funding reference 0350028.

\end{document}